\documentclass[twocolumn,english,aps,prb,floatfix,amssymb,showpacs]{revtex4}
\usepackage{amsmath}
\usepackage{amssymb}
\usepackage{graphicx}
\ifx\pdftexversion\undefined
\usepackage[dvips]{hyperref}
\else
\usepackage{hyperref}
\fi
\hypersetup{
  colorlinks = true, linkcolor = magenta
}

\newcommand{\ssm}{\scriptscriptstyle\rm}
\newcommand{\pdag}{\phantom{\dag}}
\renewcommand{\theta}{\vartheta}

\begin{document}

\title{Geometry induced pair condensation}
\begin{abstract}
We study a one-dimensional model of interacting bosons on a lattice with two flat bands. Regular condensation is suppressed due to the absence of a well defined minimum in the single particle spectrum. We find that interactions stabilize a number of non-trivial phases like a pair (quasi-) condensate, a supersolid at incommensurable fillings and valence bond crystals at commensurability. We support our analytical calculations with numerical simulations using the density matrix renormalization group technique. Implications for cold-atoms and extensions to higher dimensions are discussed.
\end{abstract}

\date{\today}

\author{Murad Tovmasyan}
\author{Evert~P.~L.~van\ Nieuwenburg}
\author{Sebastian~D.\ Huber}
\affiliation{Institut f\"ur Theoretische Physik, ETH Zurich, CH-8093 Z\"urich, Switzerland}


\maketitle

\begin{figure}[b]
\includegraphics{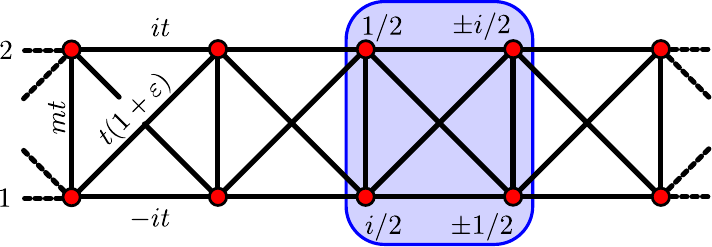}
\caption{
(Color online) {\bf Creutz ladder.} Illustration of the hopping amplitudes on the cross-linked Creutz ladder. The hopping amplitudes along the legs are such that the particles pick up a phase $\pi$ when going around a square plaquette. The blue (gray) box indicates the Wannier function in the flat band limit for the two bands ($\pm$); the numbers correspond to the respective amplitudes.
}
\label{fig:ladder}
\end{figure}

\emph{Introduction.}$-$Weakly interacting bosons at zero temperature form a Bose Einstein condensate. This condensation is an ubiquitous phenomenon and describes a large variety of physical systems such as ultra-cold atoms, short coherence-length superconductors, or ordered magnetic systems. An important avenue towards even more exciting physics is to suppress this condensation via interaction effects. The prime example is the Mott insulator where particles are localized to single sites by strong local interactions.\cite{Mott49,fisher89} While the Mott insulator is adiabatically connected to a classical state devoid of any fluctuations, it nevertheless gives rise to rich and poorly understood physics in its vicinity.\cite{Lee06,Sheng09} In this manuscript we address the question whether one can suppress condensation in a way in which already the resulting ground state is non-trivial. Moreover, we want to understand what physics can be expected close to such a state.

We investigate how the condensation of lattice bosons is suppressed due to frustrated hopping: For generic hopping problems, the long-wavelength part of the dispersion relation is quadratic and hence the lattice is rendered essentially irrelevant for the description of the low-energy physics. There is a special class of lattices, however, where interference effects due to geometric frustration lead to a low-energy behavior which is profoundly different.\cite{Moessner01b,Mila11,Huber10} Instead of possessing a quadratic minimum, the kinetic energy, $\hbar \omega(k)$, is {\em flat}, i.e., it does not depend on momentum $k$. Maybe the most striking consequence is that the group velocity $v=\partial_k\omega(k)$ vanishes for all $k$. Hence, if there is any transport through the system, it necessarily has to be due to interaction effects. We are interested in the central question if {\em repulsive contact interactions} can lead to such a mobility. More precisely, we ask for the nature of the delocalized objects. Are they dressed single particles or does the flat band give rise to more exotic physics of {\em stable} repulsively bound pairs?

Repulsively bound pairs have been observed in cold atomic gases\cite{Winkler06,Fukuhara13} and non-linear optical systems.\cite{Lahini12} The reason for their stability is simple. Two particles on the same site cost an interaction energy $U$. When separating the particles, this energy has to be converted into kinetic energy. On a lattice this might be impossible due to the finite bandwidth.\cite{Maldague77} However, such pairs are only stable if they are isolated. At a finite density, scattering of multiple pairs generically leads to their destruction. 

In this communication we show that for a flat-band system such pairs can be stabilized also at a finite density. We consider repulsively interacting bosons on a concrete one-dimensional ladder shown in Fig.~\ref{fig:ladder}. Our key result is a thermodynamically stable phase of bound pairs which (quasi-) condense, while single particle excitations are gapped. This phase is stabilized in the flat band limit. However, we show explicitly that it has a finite support away from this singular limit. This stability is crucial, both for (imperfect) experimental implementations of the model in Fig.~\ref{fig:ladder} as well as for the prospect of generalizing this phase to higher dimensions. Before going into the details of our work, we mention that such pair-condensation has attracted significant recent interest, due to its relevance for spin systems \cite{Schmidt06,Bendjama05,Momoi00} as well as for the speculated charge-$4e$ superconductor.\cite{Berg09,Paramekanti04}

\emph{Creutz ladder.}$-$We study a one-dimensional model of two cross-linked chains as depicted in Fig.~\ref{fig:ladder}. The hopping matrix elements between different legs have strength $mt$ and $(1+\epsilon)t$ for the rungs and cross-links respectively; we assume $\epsilon \geq 0$. The hopping along the legs is accompanied with an Aharonov-Bohm phase of $\pi/2$ corresponding to a $\pi$-flux through each square plaquette. The non-interacting part of the Hamiltonian is most easily written with the help of Pauli matrices $\sigma_\alpha$ encoding the two legs $\alpha=1,2$, cf. Fig~\ref{fig:ladder}. Using the Bloch operators per leg  $b_{k\alpha}^\dag$ we find
\begin{align}
H_0&=\sum_k b_{k\alpha}^\dag [\vec d(k) \cdot \vec \sigma ]_{\alpha\beta}^{\pdag}  b_{k\beta}^{\pdag},\\
\vec d &= 2t \left[\frac{m}{2}+(1+\epsilon)\cos(k),0,\sin(k)\right],
\end{align}
where we set the lattice constant $a=1$. The resulting dispersion $\hbar\omega_\pm(k)=\pm |\vec d(k)|$ has {\em two flat bands} for $\epsilon=m=0$. 

For the discussion of the interacting problem below, we need a local basis in the low-energy band $\hbar\omega_-(k)$. Wannier states form a convenient local basis for interacting flat band systems.\cite{Kohn73,Huber10} We construct them from the Bloch eigenstates
\begin{equation*}
\begin{bmatrix}
\beta_{k+}\\
\beta_{k-}\\ 
\end{bmatrix}
=
\begin{bmatrix}
\cos \frac{\theta_k}{2}  & \sin \frac{\theta_k}{2}  \\
-\sin \frac{\theta_k}{2} & \cos \frac{\theta_k}{2}
\end{bmatrix}
\begin{bmatrix}
b_{k1}\\
b_{k2}\\ 
\end{bmatrix}, \quad
\theta_k = \arctan \frac{d_x(k)}{d_z(k)},
\end{equation*}
where $b_{k\alpha}$ denote $\beta_{k\pm}$ the eigen-operators for the two bands. The Wannier states (of the lower band) are now given by 
\begin{equation}
\label{eqn:wannier}
w_{i}^\dag = \sum_j  W_1(r_i-r_j) b_{j1}^\dag + W_2(r_i-r_j) b_{j2}^\dag,
\end{equation}
with 
\begin{align}
W_1(r_i) &= 
\int_{-\pi}^{\pi} \frac{dk}{2\pi} 
e^{ik(r_i+1/2)}\sin \frac{\theta_k}{2}, \\
W_2(r_i) &= 
\int_{-\pi}^{\pi} \frac{dk}{2\pi} 
e^{ik(r_i+1/2)}\cos \frac{\theta_k}{2}.
\end{align}
Note that with the offset of $1/2$ in $\exp[ik(r_i+1/2)]$, the Wannier states are centered on plaquettes and fall off exponentially. Moreover, in the flat band limit [$m=\epsilon=0$] the Wannier states are strictly localized to single plaquettes as indicated by the blue box in Fig.~\ref{fig:ladder}. With the Wannier states (\ref{eqn:wannier}) at hand we are now in the position to tackle the interacting problem.

\emph{Effective Hamiltonian.}$-$We consider interactions in the form of a local Hubbard repulsion
\begin{equation}
H = H_0 + U \sum_{i,\alpha=1,2} b_{i\alpha}^\dag b_{i\alpha}^{\dag} b_{i\alpha}^{\pdag} b_{i\alpha}^{\pdag}.
\end{equation}
Utilizing the Wannier functions derived above we project the interaction onto the lower band ($-$). As we focus on the flat band limit, we use the Wannier functions at $m=\epsilon=0$ for the projection. Deviations from $m=\epsilon=0$ can be projected likewise and the resulting effective Hamiltonian reads
\begin{multline}
\label{eqn:heff}
H_{\ssm eff} = 
\sum_{i} \bigg[\frac{U}{4}
\rho_i(\rho_i-1) 
+\frac{U}{2} 
\rho_i\rho_{i+1} 
-\bigg(\frac{U}{8} w_i^\dag w_i^\dag w_{i + 1}^{\pdag} w_{i + 1}^{\pdag}
\\
+\frac{mt}{2} w_i^\dag w_{i+ 1}^{\pdag}
+\frac{\epsilon t}{2}  w_i^\dag w_{i+ 2}^{\pdag}
+{\rm H.c}\bigg)\bigg].
\end{multline}
Here, the index $i$ runs over a simple one-dimensional chain (of plaquette-centered Wannier operators) and $\rho_i=w_i^\dag w_i^{\pdag}$ is the number-operator on plaquette $i$. Unless stated otherwise we use the term ``site'' to describe a single ``plaquette'' in the following. However, we measure the density $n$ in the full model, e.g., we call  $n=1/4$ what corresponds to half filling in the projected model. Let us now discuss the individual terms in $H_{\ssm eff}$.

Neighboring plaquettes share two sites of the original lattice. Consequently, the local interaction can mediate both on-site as well as nearest-neighbor interactions in the projected model. These are the first two terms in (\ref{eqn:heff}). Additionally, there is a process where the interaction leads to to an effective  {\em pair-hopping} of two particles to their neighboring site [third term in (\ref{eqn:heff})].  Finally, deviations from the flat band limit in the form of non-zero $m$ and $\epsilon$ lead to the expected single particle hopping terms. Note, however, that the cross-link hopping $\propto\epsilon$ causes next-to-nearest neighbor hopping only. Before we turn to the numerical simulation of $H_{\ssm eff}$, we discuss various limiting cases where we can make definite analytical statements.
\begin{figure}[b]
\includegraphics{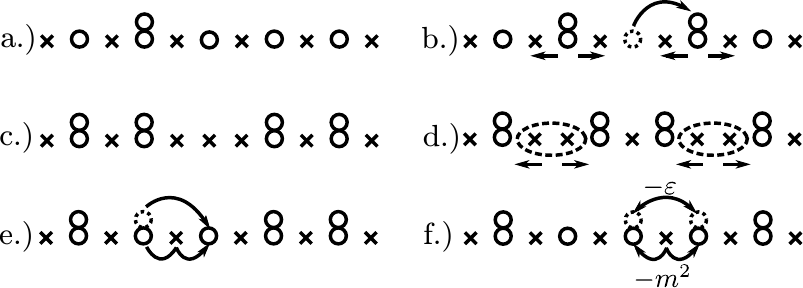}
\caption{
{\bf Processes.} Illustration of the microscopic processes responsible for the different phases. Crosses stand for empty sites, circles denote a particle. See text for a detailed description.
}
\label{fig:processes}
\end{figure}

\emph{Simple limits.}$-$In the flat band limit, the effective Hamiltonian contains only site-diagonal and {\em pair-hopping} terms . Accordingly, any static particle configuration with maximally one particle per site is an exact eigenstate of $H_{\ssm eff}$. Moreover, any such configuration with at least one empty site between any two particles is an exact zero-energy ground state of the many-body system. This leads to an exponential ground-state degeneracy below the densest packing at $n=1/4$. In the effective model this densest packing corresponds to a charge density wave. However, it is straight forward to see that in the original model each site has equal filing of $1/4$, i.e., the charge distribution is uniform and completely featureless.\cite{Kimchi13} Only the bond expectation values $\langle b_{i\alpha}^\dag b_{i+1\beta}^{\pdag} \rangle$ are modulated.  Therefore, we dub this phase a valence bond crystal (VBC).

Let us now discuss the physics arising when we dope the VBC with particles. We can put an additional particle in-between the already filled sites. This involves  a cost of twice the nearest neighbor interaction, i.e., $\delta E = U$. Moreover, there are no doubly occupied sites such that this is an exact eigenstate of $H_{\ssm eff}$ at $m=\epsilon=0$. Alternatively one can put the particle on an already filled site, cf. Fig.~\ref{fig:processes}(a). Surprisingly, this comes at half the cost regarding the effective interaction, see Eq.~(\ref{eqn:heff}). Furthermore, there is now also a doubly occupied site and hence the system can further lower its energy by delocalizing this pair over its immediately neighboring sites. Simple considerations show, that in this situation it is actually profitable to create another pair by moving two neighboring particles in the VBC on top of each other, cf. Fig.~\ref{fig:processes}(b). The resulting two-pair cluster constitutes the new ground state. We expect the system to phase separate into a high and a low density region as further added particles tend to stick to this already present cluster. To deepen our understanding of the thermodynamic phases for $n>1/4$, we try to approach the system from another commensurate filling, i.e., $n=1/2$. 

At half filling a VBC of pairs is stabilized. Above, we argued that it is profitable to put additional particles on already occupied sites. Consequently, at $n=1/2$ the system prefers a configuration where every other site is doubly occupied. If we now dope this VBC with holes [see Fig.~\ref{fig:processes}(c)] it is natural to expect a standard {\em commensurate--incommensurate transition},\cite{Pokrovsky79,Schulz80} where pairs play the role of particles.  We call this phase a solitonic pair liquid (SPL) as the mobile entities are domain walls of the pair VBC, cf. Fig.~\ref{fig:processes}(d).

\emph{Numerical results.}$-$We now turn to a numerical solution of $H_{\ssm eff}$ to check if we indeed find an SPL for densities $n \leq 1/2$. To identify the different phases we look at three different correlation functions. First, the single particle Green's function
\begin{equation}
G(i)=\langle w_i^\dag w_0^{\pdag} \rangle
\end{equation}
reveals information about the presence of a ``regular'' quasi condensate. It falls off exponentially if single particle excitations are gapped. An algebraic decays signals a quasi condensate.\cite{Giamarchi04} Second, we consider the pair correlation function
\begin{equation}
P(i) = \langle w_i^\dag w_i^\dag w_0^{\pdag} w_0^{\pdag}\rangle.
\end{equation}
An algebraic decay of $P(i)$ in combination with an exponentially decaying $G(i)$ would indicate a pair quasi condensate. Finally, we examine the structure factor 
\begin{equation}
S(q) =\sum_i   \langle \rho_i \rho_0 \rangle e^{iqr_i}.
\end{equation}
For a long-range ordered (pair) VBC $S(q)$ is expected to have $\delta$-function peaks at $q=\pi$, whereas a solitonic liquid is expected to have power-law divergencies at $q=\pi \pm (n-1/2)$.\cite{Pokrovsky79,Schulz80}
\begin{figure}[!ht]
\includegraphics{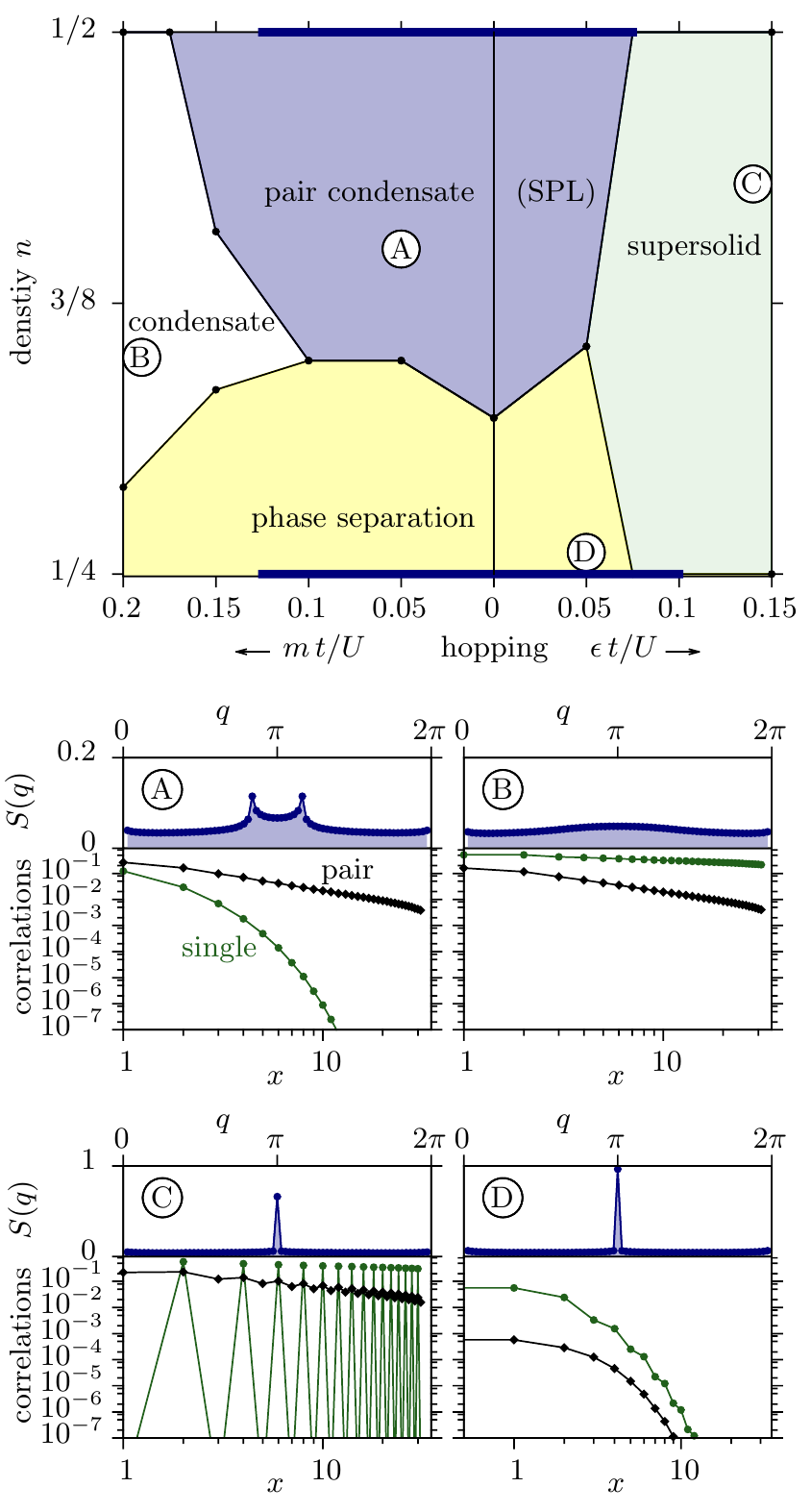}
\caption{
(Color online) {\bf Phase diagram.} (Top panel) Phase diagram as a function of the density $n$, the hopping amplitudes $m$, and $\epsilon$. The bars at $n=1/4$ and $n=1/2$ indicate the valence bond crystal (VBC) and the pair VBC, respectively. The circles mark the points in the phase diagram where we display the correlation functions in the bottom panels. (Bottom panels) Correlation functions at the respective points (A--D) in the phase diagram. On top: The structure factor $S(q)$ [normalized to $S(0)$] indicating either a solitonic liquid with two power-law peaks (A), perfect long-range order with a delta-function peaks at $q=\pi$ (C \& D), or the absence of density order in B. The Green's function $G(i)$ (green circles) and the pair correlation function $C(i)$ (black diamonds) differentiate between a pair liquid (A), a regular (quasi-) condensate (B) a super-solid (C), and the gapped crystalline phase (D).
}
\label{fig:pd}
\end{figure}

We use density matrix renormalization group (DMRG) simulations of the effective model (\ref{eqn:heff}) on chains of length up to $L=75$. We typically keep $l=400$ states and truncate the Hilbert space at a local filling $\rho_{\ssm max}=4$. We checked convergence of the results with respect to $L$, $l$, and $\rho_{\ssm max}$. 

Figure~\ref{fig:pd} summarizes our findings. The top panel of Fig.~\ref{fig:pd} shows a phase diagram as a function of $n$, $m$, and $\epsilon$. We find that the VBC ($n=1/4$) as well as the pair VBC ($n=1/2$) are stabilized also for finite $\epsilon$ and $m$, indicating a finite excitation gap in the VBC's. As expected we find a region of phase separation above $n=1/4$. We identify its extent by a negative value of the inverse  compressibility
\begin{equation}
\frac{1}{\kappa} = n^2 \frac{d^2}{d^2 n}\frac{E}{L},
\end{equation}
where $E$ denotes the ground state energy. 

For a large enough density there is a region where a SPL is stabilized. We numerically determine the Green's function $G(i)$, the pair correlation function $P(i)$, and the structure factor $S(q)$ at representative points indicated in the phase diagram, cf. Fig.~\ref{fig:pd}. Panels A and and D indeed confirm our expectation regarding the VBC and the SPL. For large enough $m$ [panel B], we find a regular (quasi-) condensate as expected for a curved band: Both the Green's function and the pair correlation function behave as a power-law while the structure factor is essentially featureless.

A finite next-to-nearest neighbor hopping $\epsilon$ stabilizes a supersolid, cf.~panel~C: Both $G(i)$ and $P(i)$ behave as power-laws and are modulated with $\exp(i\pi r_i)$. While $G(2i+1)\equiv 0$, the pair correlation function is only slightly suppressed at odd separations. Moreover, the structure factor $S(q)$ has a sharp peak at $q=\pi$, indicating true long range order in the density modulation. What process can stabilize such a long range order in a one dimensional system at incommensurate filling?

\emph{Toy model.}$-$We construct a toy model to explain the supersolid at incommensurate densities. Starting from the pair VBC we remove one pair from the lattice, cf. Fig.~\ref{fig:processes}(c). We saw that for the SPL, the array of three adjacent empty sites splits into two domain walls, cf. Fig~\ref{fig:processes}(d). This allows the system to lower its energy by delocalizing two independent domain walls, thereby destroying the long-range order in $S(q)$. The energy of such a state can be estimated to be
\begin{equation}
\frac{E_{\ssm SPL}}{U} \approx \frac{1}{4} - \frac{5}{2}\left(\frac{1}{2}-n\right).
\end{equation}

In order to stabilize the supersolid we need a mechanism to bind the two domain walls together. By breaking one pair adjacent to the hole in the pair VBC we fill all sites (of the original VBC) with at least one particle, cf. Fig.~\ref{fig:processes}(e). In the presence of single particle hopping we can now delocalize these half-empty sites. However, the effective hopping amplitude is different for nearest neighbor hopping $m$ and next-to-nearest neighbor hopping $\epsilon$. The latter can hop particles resonantly on sites that are filled with one particle, cf. Fig.~\ref{fig:processes}(f). The former, on the other hand, has to hop via an intermediate state which is off-resonant by $\delta E = U/2$. Therefore, we can estimate the energy for a supersolid where single-particles are delocalized as
\begin{equation}
\frac{E_{\ssm SS}}{U} = \frac{1}{4} -\frac{3}{2}(1-n) 
-4(1-n) \frac{t}{U} \times
\begin{cases}
\epsilon \\
m^2
\end{cases}.
\end{equation}

A third option is to condense the particles into a regular condensate. We get a simple estimate of the energy by just replacing $w_i\rightarrow \sqrt{n}$
\begin{equation}
\frac{E_{\ssm BEC}}{U}  = \frac{3}{4} n^2 - \frac{1}{4} n - n \frac{t}{U}  \times
\begin{cases}
\epsilon\\
m
\end{cases}.
\end{equation}
The estimates of $E_{\ssm SPL}$ and $E_{\ssm SS}$ are only valid close to $n=1/2$ and for $m,\epsilon \ll 1$. Comparing the energies close to this filling we see that domain walls can be bound for $\epsilon > 0$ and hence a supersolid is stabilized. For $m>0$, the regular condensate $E_{\ssm BEC}$ always wins over the supersolid, a finding that is in accordance with the DMRG results.

\emph{Discussion.}$-$We studied a one-dimensional ladder with two perfectly flat Bloch bands. A local Hubbard interaction leads to the stabilization of various localized phases at commensurable fillings. Away from commensurability the system undergoes either phase separation or exhibits a pair (quasi-) condensate. This condensate melts into a supersolid via a binding mechanism for the freely moving domain walls. Alternatively a regular uniform (quasi-) condensate is stabilized. By finding a pair condensate with a finite extent in the $m$--$\epsilon$ plane we can answer the question posed in the introduction affirmatively: We identify a system where local repulsion can lead to stable pairs in a many-body context. An interesting way to check our predictions in an experiment would be via transport measurement in cold atoms as pioneered in Ref.~\onlinecite{Brantut12}.

\emph{Acknowledgements.}$-$We acknowledge stimulating discussions with Erez Berg, Andreas R\"uegg, Roman S\"usstrunk, and Oded Zilberberg. In the final stages of writing this manuscript we got aware of a similar study \cite{Takayoshi13} restricted to the completely flat band limit.

\end{document}